\documentstyle[aps,twocolumn,epsf]{revtex}
\begin{document}
\draft
\title{Noether Currents of Charged Spherical Black Holes}
\author{M. C. Ashworth\thanks{\tt ashworth@th.phys.titech.ac.jp}}
\address{Department of Physics, Tokyo Institute of Technology,
Ohokayama, Meguro, Tokyo 152-8551, Japan}
\author{Sean A. Hayward,\thanks{\tt hayward@gravity.phys.psu.edu}}
\address{Center for Gravitational Physics and Geometry,
104 Davey Laboratory, The Pennsylvania State University,
University Park, PA 16802-6300, U.S.A.}
\date{7th April 2000}
\maketitle

\begin{abstract}
We calculate the Noether currents and charges for Einstein-Maxwell
theory using a version of the Wald approach. In spherical
symmetry, the choice of time can be taken as the Kodama vector.
For the static case, the resulting combined Einstein-Maxwell
charge is just the mass of the black hole. Using either a
classically defined entropy or the Iyer-Wald selection rules, the
entropy is found to be just a quarter of the area of the trapping
horizon. We propose identifying the combined Noether charge as an
energy associated with the Kodama time. For the extremal black
hole case, we discuss the problem of Wald's rescaling of the
surface gravity to define the entropy.
\end{abstract}

\pacs{}

It is widely excepted that black holes have entropy. However,
without a full quantum theory of gravity, the statistical origin
of this entropy is still unclear. On a classical and
semi-classical level there have been many proposals discussing
where this entropy comes from and ways of calculating it: spin
structures, entanglements, edge states, etc. In particular
interest for this paper is the Noether current calculation of Wald
and Iyer \cite{Wald1}- \cite{Iyer2}. The question becomes one of
how do we test these proposals with our current level of
understanding of gravity and quantum gravity?  To date most of
these calculations have been done in static or in some sense
quasi-static cases \cite{static}.  However, such cases may not
reflect the general nature of the black hole. What is needed is
further test cases, in particular dynamical test cases.

In this regards, spherically symmetric space-times provides us
with a suitable dynamical testing ground.  In the spherically
symmetric case, while remaining dynamical, we can identify certain
features defined in static cases. The time-like Killing vector
that is identified as time in the static case can be replaced with
the Kodama vector, $k = {\star} d r$ where $r=\sqrt{{\mathcal
A}/4\pi}$ is the areal radius and $\star$ is the Hodge operator of
the 2D normal space\cite{SH1}. In addition, a local active
gravitational energy can be defined by the Misner-Sharp energy
\cite{Miser},
\begin{equation}
E = {r \over 2} ( 1- d r \cdot d r)
\end{equation}
For dynamical space-times, a locally defined horizon and the
global event horizon are generally not the same. Therefore, we
must choose a suitable definition of the outer surface of the
black hole. In the spherically symmetric setting, the outer
surface of the black hole is proposed to be the trapping horizon
defined by $\nabla r$ being null everywhere on the horizon
\cite{SH1}.  In such a case, the energy on the trapping horizon
$E$ is just half the areal radius, a natural generalization of the
Schwarzschild radius.

Identifying these properties of the black-hole space-time, it is
possible to study the thermodynamics in a classical setting. The
equations of motion can be shown to give an energy balance
equation \cite{SH1},
\begin{equation}
\nabla E = {\cal A} \psi + w \nabla {\cal V}
\end{equation}
where $\psi$ is a localization of the Bondi energy flux, $w$ is an
energy density and ${\cal V}={4\over3}\pi r^3$ is the areal
volume. Looking more closely, we can identify the second term on
the right-hand side as a work term. The first term on the
right-hand side is an energy supply. This term, again using the
equations of motion, can then be written as
\begin{equation} \label{enflux1}
{\cal A} \psi = { \kappa \nabla {\cal A} \over 8 \pi} + r \nabla
\left( E \over r \right),
\end{equation}
where the dynamical surface gravity is defined as for stationary
surface gravity by replacing the Killing vector with the Kodama
vector \cite{SH1}, yielding  $\kappa = {\star} d k / 2$.  The last
term vanishes when projected along a trapping horizon. If
$\kappa/2\pi$ is the temperature on the trapping horizon, then the
entropy is given by the area of the trapping horizon as ${\mathcal
A} / 4$.

So, in a spherically symmetric system, we can identify the state
variables of the model.  The kinematical quantities are given by
the areal radius and dynamical time $(r,k)$. The gravitational and
matter quantities are $(E,\kappa)$ and $(w,\psi)$ respectively.
Normally because there is no preferred time, such quantities are
difficult to define.

In this testing ground, we would like to look at the Noether
current calculations of Wald and Iyer in more detail.  In general,
gravitational theories are defined from diffeomorphism invariant
actions, specified by a Lagrangian $n$-form $L[\phi]$, where
$\phi$ denotes the dynamical fields including the space-time
metric. For every such Lagrangian, there is an associated
conserved current and charge, as follows, simplifying Wald's
method by considering only perturbations $\delta$ which are Lie
derivatives ${\cal L}_\xi$ along a vector $\xi$, which is the
local generator of the diffeomorphisms. Then
\begin{equation}
\delta L={*}\Phi\circ\delta\phi+d\Theta
\end{equation}
defines the boundary $(n-1)$-form $\Theta[\phi,\xi]$, where $*$ is the
space-time Hodge operator. The bulk term $\Phi$, which gives the
equations of motion, has tensorial indices dual to $\phi$, with
$\circ$ denoting contraction of all indices. This leads to a
current $(n-1)$-form
\begin{equation}
J = \Theta - \xi \cdot L.
\end{equation}
Then the identity
\begin{equation}
{\cal L}_\xi\Lambda=\xi\cdot d\Lambda+d(\xi\cdot\Lambda)
\end{equation}
implies
\begin{equation}
dJ=-{*}\Phi\circ{\cal L}_\xi\phi
\end{equation}
which vanishes when the equations of motion hold, $\Phi=0$. On
shell $J$ is closed, and
\begin{equation}
J=dQ
\end{equation}
defines a conserved charge $(n-2)$-form $Q$, up to various gauge
freedom.

Originally for the Einstein action, Iyer and Wald \cite{Iyer1}
found that integrating the charge associated with the Killing
time, after rescaling the surface gravity to one over the
bifurcation surface, gave the known entropy of the static black
hole. In the spherical setting, we can replace the Killing time
with the Kodama vector as the diffeomorphism generator in the time
direction. Integrating the charge from the Einstein action $L_E=
\ast R / 16 \pi$ over a section of the trapping horizon instead of
the bifurcation surface, we get the same form \cite{SH2}
\begin{equation}
\oint Q_{E} = {{\cal A} \kappa \over 8 \pi}. \label{charge1}
\end{equation}
In the static case when $k$ commutes with $d r$ the Kodama vector
reduces to the Killing time \cite{SH1}, recovering the Wald-Iyer
result; an entropy that is just the area of the horizon.  However,
we have now moved off the bifurcation surface to the locally
defined trapping horizon.\footnote{Note that in Jacobson et al.
\cite{JKM1} showed that any section of a Killing horizon is
equivalent to the bifurcation surface.}

In general, $\kappa$ is dynamical, the surface temperature of the
black hole varying with time as the black hole area changes, so
such rescaling of the charge seems somewhat artificial, as pointed out
in \cite{SH2}. So the question that then comes to mind is: what
is this conserved current and charge? To get a broader look at
this question, we should add matter fields to that action. So, let
us consider the case that the matter fields are given by the
Maxwell electromagnetic action,
\begin{equation}
L_M[g,A] = - {1 \over 16 \pi} \ast F : F,
\end{equation}
where the 1-form $A$ is the electromagnetic potential,
$F=2dA$ and $:$ is the trace of the dot product. Then we
find
\begin{equation}
\delta L_M={*}(T:\delta g/2+\Psi\cdot\delta A)+d\Theta,
\end{equation}
where
\begin{eqnarray}
&&T=-(F\cdot F+(F:F)g/4)/4\pi\\ &&\Psi={*}d{*}F/4\pi\\ &&\Theta  =
- \delta ( {*} A \cdot F )/8\pi.
\end{eqnarray}
The resulting Noether current is
\begin{equation}
J_M = d Q_M - \xi\cdot{*} ( \Psi \cdot A )/2,
\end{equation}
where the associated charge is
\begin{equation}
Q_M = - {1 \over 8 \pi } \xi \cdot \ast (A \cdot F).
\end{equation}
Note that this method differs from that of Wald in that it has not
been necessary to express $\Theta$ explicitly as a function of
$(\phi,\delta\phi)$, thereby saving calculation. The calculation
also uses the fact that $d$ and ${\cal L}_\xi$ commute.  Note:
this charge is dependent on the gauge choice of the EM fields,
which is related to the Aharonov-Bohm effect.

We propose defining the energy of a field as twice the charge $Q$
integrated over an $(n-2)$-dimensional surface. For the
Reissner-Nordstr\"om case, with the natural gauge choice $A=-e \ d t
/r$, this gives the electromagnetic energy
\begin{equation}
E_M= \oint 2 Q_M =  {e^2 \over r}
\end{equation}
Combining this with the twice the Noether charge from the Einstein
gravitational action (\ref{charge1}),
\begin{equation}
E_{E}= \oint2 Q_{E} = {{\cal A}\kappa\over{4\pi}},
\end{equation}
gives a combined energy charge,
\begin{equation}
E = E_{E} + E_M = m,
\end{equation}
that is just the mass of the black hole.

In \cite{MASH1}, the question of the physical meaning of
 the total Noether current is
brought up. The charge normally associated with transformations
along a time direction is of course the energy.  Although
further cases need to be studied, in the Schwarzschild case and
now the Reissner-Nordstr\"om case, the total Noether charge is
indeed the energy at
the boundary.

Another point about the Noether currents is that because the
symplectic potential and the charge are only defined up to total
derivatives, it is possible to change the charge by adding various
surface terms.  Iyer and Wald have proposed methods for choosing
which terms should lead to the entropy \cite{Iyer2}, which
does give the right entropy in the Reissner-Nordstr\"om
case after rescaling the surface gravity. However, it seems easier
to understand the freedom of the choice of the symplectic
potential if the Noether charge is viewed as an energy. The
changes in the current are analogous to Legendre transformations
in thermodynamics which result in different energies such as the Gibb's
free energy. So with Iyer and Wald's
selection rule \cite{Iyer2}, the gravitational entropy seems to
be singled out such that it occurs with only the surface gravity
appearing in front of it.

In the above calculations, it is assumed that the space-time is
the normal black hole space-time resulting from $e^2 < m^2$.  In
the case that $e^2 > m^2$, there is no trapping horizon and no
other inner boundary other than the singularity at $r=0$. In the
extremal case $e^2 =m^2$, the inner and outer trapping horizons
coincide and become degenerate. Using standard null coordinates
\cite{Hawking},
\begin{equation}
ds^2  =   - \left( {(r-m)^2 \over r^2 } \right) dx^+ dx^- + r^2 d
\Omega^2
\end{equation}
where the areal radius is defined indirectly by
\begin{equation}
{1 \over 2} (x^+ - x^-)   =  r + 2 m \log (r-m) - { 2 m^2 \over
r-m} \\[1ex]
\end{equation}
it is easy to see that $r=m$ is a still a well defined trapping
horizon.  However, the surface gravity is proportional to the
difference between the inner and outer trapping horizon radius,
$\kappa \propto (r_+ - r_-)$, and is zero in the extremal case. In
Wald's definition, the surface gravity must be rescaled to unity
in order to define the entropy.  In the extremal case, because the
surface gravity is zero, it is not possible to do this rescaling,
resulting in an ill-defined entropy. Using the classical
first law (\ref{enflux1}), it would still seem that the
entropy is just the area of the trapping horizon over 4. However, using
Nernst's theorem: if the temperature (the surface gravity)
vanishes then the system settles in its ground state and the
entropy vanishes. The system being in some sort ground state makes
sense because the Einstein energy vanishes, $E_{E}=0$. The problem
of the entropy of the extremal case has been discussed in various
papers, for example \cite{HHR}. However without a
quantum-statistical model for the entropy, this problem cannot be
resolved.

\bigskip
Acknowledgements. MCA is supported by NSF/JSPS Postdoctoral
Fellowship for Foreign Researchers (No.\ P97198). SAH is supported
by National Science Foundation award PHY-9800973.



\begin{references}
\bibitem{Wald1} Robert M. Wald, Phys. Rev. D {\bf 48},
3427-3431 (1993).

\bibitem{Iyer1} Vivek Iyer, Robert M. Wald, Phys.Rev. D {\bf 52},
4430-4439 (1995).

\bibitem{Iyer2} Vivek Iyer and Robert M. Wald, Phys.Rev. D {\bf 50},
846-864 (1994).

\bibitem{static}
A. Ashtekar, J. Baez, A. Corichi, and K. Krasnov, {Phys. Rev.
Lett.} {\bf 80}, 904 (1998); J. Gegenberg, G. Kunstatter, and D.
Louis-Martinez, Phys. Rev. {\bf D51}, 1781 (1995); S. Carlip,
Class. Quantum Grav. {\bf 12}, 2853 (1995); S. Carlip,
Phys.Rev.Lett. 82, 2828-2831 (1999).

\bibitem{SH1} Sean Hayward, Class.Quant.Grav. {\bf 15},
3147-3162 (1998).

\bibitem{Miser} C. W. Misner and D. H. Sharp, Phys. Rev.  B {\bf
571}, 136 (1964).

\bibitem{SH2}
S. A. Hayward, S. Mukohyama, M. C. Ashworth, Phys.Lett. A256,
347-350 (1999).

\bibitem{JKM1} T. Jacobson, G. Kang, R. C. Myers,
Phys. Rev. {\bf D49} 6587-6598 (1994).

\bibitem{MASH1} M. C. Ashworth and S. A. Hayward,
Phys.Rev. D60, 084004 (1999).

\bibitem{Hawking} S. W. Hawking and G. F. R. Ellis, {\it The Large
Scale Structure of Space-time} (Cambridge University Press, 1973).

\bibitem{HHR} S. W. Hawking, G. T. Horowitz, and S. F. Ross,
Phys.Rev. {\bf D51}  4302-4314 (1995).

\end{references}
\end{document}